\documentclass[conference]{IEEEtran}

\usepackage{cite}

   \usepackage{graphicx}

\usepackage{textcomp}

\begin{document}
%
\title{Superconductivity in the Iron-Pnictide Parent Compound SrFe$_2$As$_2$}

\author{\IEEEauthorblockN{Kevin C. Kirshenbaum, Shanta R. Saha, Nicholas P. Butch, \\Jeffrey D. Magill, and Johnpierre Paglione}
\IEEEauthorblockA{Center for Nanophysics and Advanced Materials\\
University of Maryland\\
College Park, MD 20783\\
Email: paglione@umd.edu}}

\maketitle

\begin{abstract}
The appearance of superconductivity at ambient pressures in the undoped iron pnictide parent compound SrFe$_2$As$_2$ is studied experimentally using several techniques and approaches to aid in understanding the nature of this phase. Low temperature magnetization measurements of single crystals of SrFe$_2$As$_2$ reveal diamagnetic screening due to the onset of superconductivity below 21 K, with volume fraction estimates varying between 0 and 15\%. The effects of heat treatment and cold-working via severe mechanical deformation on the superconducting phase found are studied, showing that superconductivity can be suppressed via modest annealing and reinstated by cold-working.
\end{abstract}


\IEEEpeerreviewmaketitle

\section{Introduction}
High temperature superconductivity research remains a relatively new and ongoing field of research.  The discovery of high temperature superconductivity in cuprates in 1986 \cite{Bednorz189} led to renewed interest in the phenomenon of superconductivity that, until then, had a transition temperature limit of around 30 K \cite{McMillan331}.  Despite much study, the pairing mechanism of high temperature superconductivity still remains a mystery and so the discovery of new classes of high temperature superconductors is a priority, offering physicists new perspectives from which to study this fascinating phase of matter.
 
Since the discovery of quaternary pnictide superconductors in early 2008 \cite{Kamihara3296} three other types of crystal structures containing iron and pnictogen group elements have been found to exhibit similar magnetic and superconducting properties.  These are the binary FeSe systems, tertiary \(A\)FeAs (\(A\) = alkali metal), and the 1-2-2 system \(AE\)Fe$_2$As$_2$ (\(AE\) = alkaline earth metal) \cite{Rotter020503}.  Of these, the highest value of $T_c$ obtained thus far has been reported at 55 K in the quaternary system \(Ln\)FeAsO (\(Ln\) = rare earth element) \cite{Ren2215}.  With a slightly lower maximum value of $T_c$ (38 K in Ba$_{1-x}$K$_x$Fe$_2$As$_2$ \cite{Rotter107006}), the first large single crystals were obtained in the intermetallic 1-2-2 system, and have therefore made this a convenient system to study.  

As is well known with the cuprate and heavy-fermion superconductor families, superconductivity in these materials is often established only after tuning the system via doping or pressure \cite{Stewart755, Butch106, Heffner361}.  Likewise, it is widely accepted that superconductivity is stabilized in the iron pnictide materials by either electron- or hole-doping of the system, or by applying an external pressure. In most pnictide superconductors, the undoped, unpressurized materials, the so-called ``parent'' compounds, do not exhibit superconductivity but rather undergo magnetic ordering like the cuprates and heavy fermion systems.  The magnetically ordered phase can be suppressed through doping or pressure, and this suppression is often accompanied by the presence of superconductivity over a limited range of parameter space \cite{Tapp060505}.  SrFe$_2$As$_2$, for example, undergoes a magnetostructural phase transition at $\sim 200$~K that can be suppressed by applied pressure or chemical substitution \cite{Tegel452201, Kotegawa013709, Kim102203, Chen224512}.

However, observations by Saha {\it et al.} of a superconducting phase in undoped, unpressurized single crystals of the parent compound SrFe$_2$As$_2$ \cite{Saha0811} have raised much interest. Furthermore, recent observations of a similar superconducting transition in thin films of undoped SrFe$_2$As$_2$ \cite{Hiramatsu0903}, and the related 1-2-2 compound BaFe$_2$As$_2$ \cite{Tanatar0903} suggest that this phase is not exclusive to the particular crystal growth method used, nor to the particular compound SrFe$_2$As$_2$ itself, but may be a more widespread phenomenon in this family of materials. 
As discussed by Saha {\it et al.}, this unexpected superconducting phase typically occupies less than 15\% of sample volumes but occurs in 90\% of grown samples and is robust with respect to applied magnetic fields \cite{Saha0811}.
Here we report an investigation of this phenomenon in SrFe$_2$As$_2$ single crystals measured at ambient pressure, describing how the superconducting phase can be removed by heat treatment and how it can be reestablished by applied mechanical stress.

\section{Growth and Characterization of Single Crystals}

SrFe$_2$As$_2$ single crystals were grown using the self-flux method described by Wang {\it et al.} \cite{Wang117005}.  A stoichiometric mixture of Fe (5N purity) and As (4N) was pre-reacted in evacuated quartz tubes and sintered at 700 \textdegree C for 24 hours.  After the FeAs was formed it was characterized via energy dispersive x-ray spectroscopy (EDS) and found to have less than 5\% impurity phases (mostly Fe$_2$As).  The sintered FeAs was mixed with elemental Sr (3N5) in a ratio of 1:4 Sr:FeAs in either an Ar atmosphere or under vacuum.  The mixture was then heated to 1200 \textdegree C, cooled at a rate of 2 \textdegree C per hour until it reached 950 \textdegree C, then allowed to cool to room temperature slowly.  Crystals have been grown both in alumina crucibles and in bare quartz tubes with no observable differences in the resultant crystals.  This process has enabled the growth of large crystals with typical dimensions 5 $\times$ 5 $\times$ 0.25 mm$^3$ essentially limited by crucible size.  Note that the flux is not decanted in this process, instead requiring simple mechanical separation of single crystals from the hardened flux matrix.

Structural properties were characterized by both powder and single-crystal x-ray diffraction using Rietfeld refinement (SHELXS-97).  The lattice was determined to have the previously reported \cite{Tegel452201} I4/mmm structure with lattice constants $a $= 3.9255(2) \AA\ and $c$ = 12.3172(12) \AA.  EDS analysis determined that the crystals have the desired 1:2:2 stoichiometry with no impurity phases detected.  Resistivity and specific heat were measured in a Quantum Design Physical Properties Measurement System using the four-probe AC method, with silver paint wire contacts made at room temperature.  Magnetic susceptibility was measured in a Quantum Design Magnetic Properties Measurement System.

\begin{figure}
	\includegraphics[width = 3.25in]{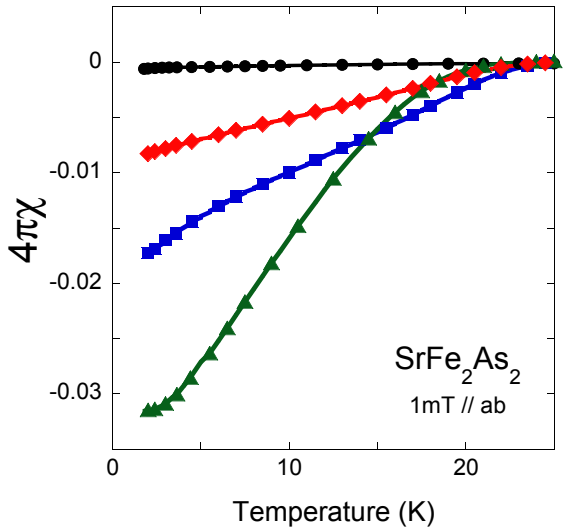}
	\caption{Diamagnetic screening in four SrFe$_2$As$_2$ single crystals as measured by magnetic susceptibility, showing the estimated superconducting volume fraction to range from zero to slightly more than three percent of the total volume of the samples. The volume fraction ranges up to $\sim 15\%$ in as-grown crystals \cite{Saha0811}.}
\end{figure}

\section{Results \& Discussion}

\subsection{Superconducting Volume Fraction and Annealing}

Figure 1 presents low temperature magnetization data taken on several as-grown crystals of SrFe$_2$As$_2$. As shown for four representative samples, the superconducting volume fraction (as estimated from the fraction of full diamagnetic screening $4\pi\chi$) varies continuously between a negligible amount to approximately 3\%. This fraction has been shown to reach as high as $\sim 15\%$ \cite{Saha0811}, suggesting that this volume fraction can reach non-negligible proportions.

As shown in Figure 2, a complete resistive transition is observed below T$_c=21$~K, which is typical observed in a majority of as-grown SrFe$_2$As$_2$ samples grown in either alumina crucibles or directly in quartz tubes. It should be noted that partial resistive transitions, where a complete zero-resistance state is not achieved, are also observed in some samples presumably due to the presence of a very small $\ll 1\%$ volume fraction. However, it appears that the transition temperature always occurs near 21~K regardless of complete or incomplete resistive transition, or of the percentage of screening.

This partial-volume superconducting phase has proven to be particularly sensitive to mild heat treatment.  In Figure 2 we show resistivity as a function of temperature for a single sample SrFe$_2$As$_2$ taken both before and after an annealing treatment (data is for the same sample but with different electrical contacts made after heat treatment). As discussed above, there is a complete resistive transition observed for the as-grown crystal, suggesting a sufficient volume of the sample is superconducting to provide a complete shorted path between sample contacts, thus shorting out the voltage measurement.
After heating the sample for only 10 minutes at 300 \textdegree C in a sealed argon environment devoid of oxygen, the superconducting transition is shown to be completely suppressed. In other experiments (not shown), it was determined that it takes as little as 200 \textdegree C to suppress this transition, with some samples still showing a partial drop or kink at the same $T_c$ value after low temperature heating. Also, annealing in air or under sealed argon atmosphere has the same effect in suppressing the superconducting transition. It may be thought that some type of impurity or oxidation phase that is changed after heat treatment may give rise to the change in superconducting properties in such samples. However, both EDS and WDS analysis has been performed on samples both before and after our annealing procedure, showing no changes to the chemical composition detected. This is unlike the scenario found in annealed cuprate superconductors, which show a drastic change in the superconducting properties associated with a reduction in oxygen present in the crystals \cite{Kang224, Jiang8151}. In contrast, there is no indication that oxygen plays any role in the phenomena reported here.

\begin{figure}
	\includegraphics[width = 3.25in]{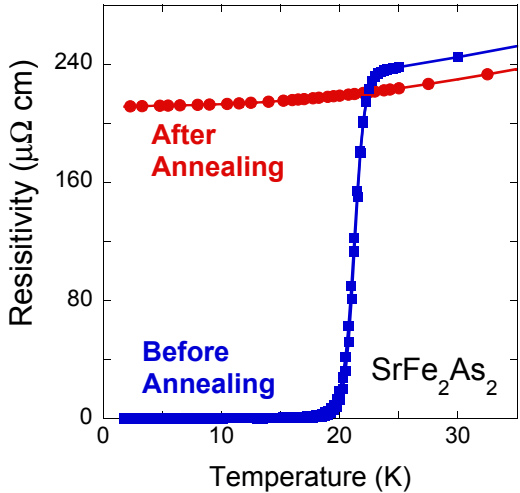}
	\caption{Resistivity of a single-crystal sample of SrFe$_2$As$_2$, measured before (blue squares) and after (red circles) a 10 minute annealing treatment at 300 \textdegree C in argon atmosphere. Data is not normalized.}
\end{figure}

\subsection{Deformation Experiment}

Interestingly, the application of mechanical stress to crystals without any trace of a resistive transition has been shown to reinstate the superconducting phase that was suppressed during annealing.  Studies by Saha {\it et al.} have shown that applying a hydrostatic pressure of 14 kbar and then releasing the pressure revives the full resistive transition \cite{Saha0811}. In order to investigate the role of mechanical stress in inducing superconductivity, several other physical tests were performed, including cleaving, quenching with liquid nitrogen, and subjecting a sample to severe mechanical deformation. 

A large as-grown single-crystal sample was first annealed to remove any trace of superconductivity and then subject to several tests. Figure 4 presents the resistivity measurements taken after annealing the sample and after each stress treatment to check for any trace of induced superconductivity. First, the sample was plunged into liquid nitrogen to test the ability of rapid thermal cycling to induce superconductivity.  However, resistivity measurements found that the superconducting transition did not return.  Next the sample was cleaved in order to investigate whether our standard procedure for preparing resistivity samples was in fact responsible for inducing superconductivity. The sample was carefully cleaved into two pieces using a steel razor blade, after which the resistivity of each half was measured.  Again, there was no return of a superconducting transition, as shown in Figure 3. Finally, a remaining piece of the sample with approximate size 1 mm$^2$ was placed in a piston-type press and squashed until its surface area had doubled, requiring an applied force of 4,000 lbs on a 1~cm diameter piston. 

In contrast to the previous tests, this severe deformation was found to induce a return of the 21 K superconducting transition. Figure 3 presents the resistivity of the squashed sample in comparison to its resistivity just after annealing and after cleaving, showing that the major difference between the three curves is in the appearance of the drop in resistivity at $T_c$ in the deformed sample.
(The resistivity data were shifted and normalized to the values of the severely deformed sample at 25 and 300 K to remove a substantial increase in the residual resistivity and change in the geometric factor of the deformed sample arising from defects, cracks, etc. introduced during the deformation process.) Although the measured drop is not a complete transition, it is presumed that the actual volume fraction of superconductivity may be larger than this drop implies, due to the possible introduction of fractures and voids in the sample that would interrupt a continuous superconducting pathway.
Unfortunately, this level of mechanical deformation weakens the sample enough that it is difficult to do subsequent characterization measurements. Work is under way to do a similar experiment but using magnetization to determine the change in the superconducting volume fraction induced by mechanical deformation of a larger crystal.

\begin{figure}
	\includegraphics[width = 3.25in]{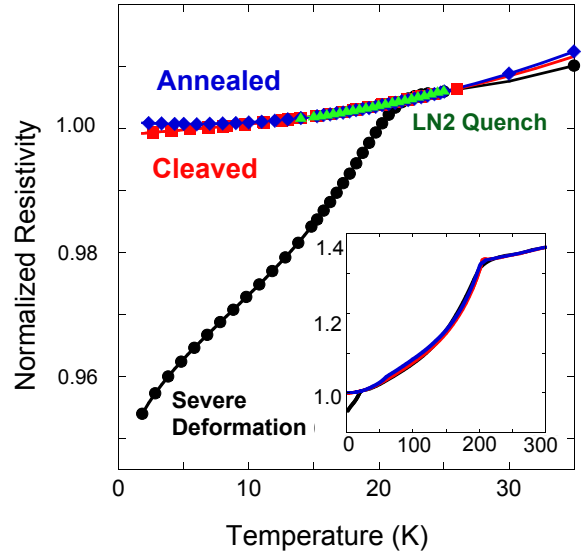}
	\caption{Electrical resistivity measurements of a single-crystal sample of SrFe$_2$As$_2$ subject to various treatments. The sample was first annealed at 300 \textdegree C to suppress any trace of superconductivity (blue diamonds), then quenched in liquid nitrogen from room temperature (green triangles), then cleaved with a razor blade (red squares), and one remaining piece subsequently pressed to $\sim 100\%$ mechanical deformation (black circles). Data is shifted and normalized to the resistivity of the pressed sample at 25 K to remove spurious contributions from micro-cracks, etc. as discussed in the text.  The inset shows the resistivity profile over the full temperature range.}
\end{figure}

Figure 4 presents scanning electron microscope images of the surface of a sample subject to the same deformation treatment, indeed showing the presence of micro-cracks throughout the sample. These cracks are presumably the reason for the absence of a complete resistive transition, although more studies are required to make such a determination. While more tests are under way to investigate the details by which stress-induced superconductivity occurs in SrFe$_2$As$_2$, a correlation between crystal lattice deformation and the superconducting phase has been shown to exist in single-crystal X-ray diffraction experiments, pointing to a scenario where superconductivity is stabilized by a type of internal strain arising from tilted stacking faults \cite{Saha0811}. It remains to be determined how this seemingly intrinsic level of distortion in pristine crystals is connected with the severe deformation effect. It would be interesting, for instance to investigate this using a more powerful probe such as transmission electron microscopy.

\begin{figure}
	\includegraphics[width = 3.25in]{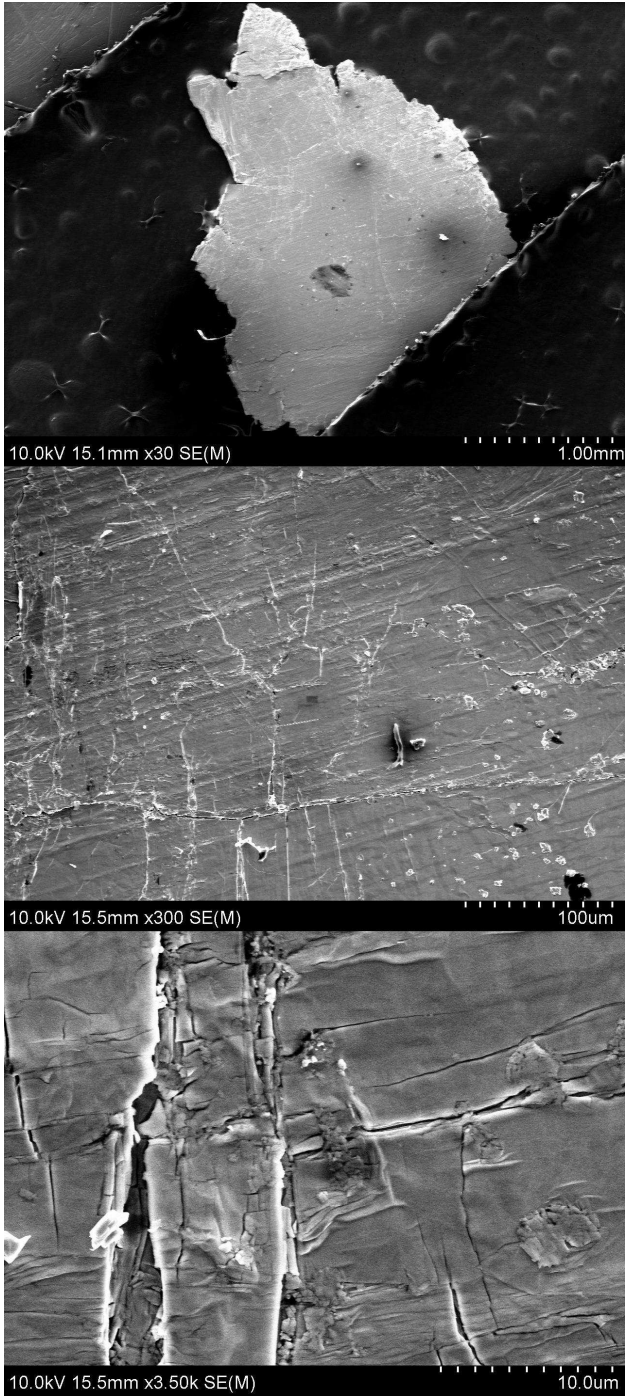}
	\caption{Scanning electron microscopy images of a single-crystal sample of SrFe$_2$As$_2$ subject to severe ($\sim 100\%$) deformation by a piston-type press. Panels a), b) and c) show pictures of the same sample at different magnification levels as indicated.}
\end{figure}

\section{Conclusion}
In conclusion, the ambient pressure superconducting phase that appears in a majority of single-crystal samples of SrFe$_2$As$_2$ synthesized using the self-flux method was studied via resistivity and magnetization measurements of samples subject to heat treatment and mechanical stress tests. The partial volume fraction superconductivity was shown to be suppressed by modest heat treatments, while it can be reinstated by subjecting the material to severe mechanical stress. Future experiments will investigate the nature of stress-induced deformation and its role in stabilizing superconductivity in this iron-pnictide parent compound.

\hfill 1 May 2009


\section*{Acknowledgment}

The authors acknowledge R. Suchoski, P. Y. Zavalij, and B. W. Eichhorn for experimental assistance, and P. Bach, K. Jin, X. Zhang, and R. L. Greene for useful discussions.  N. Butch is supported by a Glover Fellowship from the Center for Nanophysics and Advanced Materials.

\bibliographystyle{IEEEtran}

\end{document}